\documentclass[twocolumn,amssymb,showpacs, amsmath,nobibnotes,nofootinbib,aps,prb]{revtex4}
\usepackage[dvips]{graphicx}
\usepackage[dvips]{color}
\usepackage{subfigure}
\begin{document}\title{Neutron inelastic scattering investigation of the magnetic excitations in Cu$_2$Te$_2$O$_5$X$_2$ (X=Br, Cl)}
\author{S. J. Crowe}
\author{S. Majumdar}
\author{M. R. Lees}
\author{D. M$^c$K. Paul}
\affiliation{Department of Physics, University of Warwick, Coventry CV4 7AL, United Kingdom}
\author{R. I. Bewley}
\author{S. J. Levett}
\affiliation{ISIS Facility, Rutherford Appleton Laboratory, Chilton, Didcot, Oxon OX11 0QX, United Kingdom}
\author{C. Ritter}
\affiliation{Institut Laue-Langevin, 156X, 38042 Grenoble C\'edex, France}
\begin{abstract} 
Neutron inelastic scattering investigations have been performed on the spin tetrahedral system Cu$_{2}$Te$_{2}$O$_{5}$X$_{2}$ (X = Cl, Br). We report the observation of magnetic excitations with a dispersive component in both compounds, associated with the 3D incommensurate magnetic order that develops below $T^{Cl}_{N}$=18.2 K and $T^{Br}_{N}$=11.4 K. The excitation in Cu$_{2}$Te$_{2}$O$_{5}$Cl$_{2}$ softens as the temperature approaches $T^{Cl}_{N}$, leaving diffuse quasi-elastic scattering above the transition temperature. In the bromide, the excitations are present well above $T^{Br}_{N}$, which might be attributed to the presence of a degree of low dimensional correlations above $T^{Br}_{N}$ in this compound.
\end{abstract}

\pacs{61.12.Ex, 75.10.Jm, 75.10.Pq}

\maketitle 
\par
The low-lying excitations in spin triangular or tetrahedral lattice systems often lead to exotic magnetic behavior. Recently, the spin tetrahedral compounds Cu$_{2}$Te$_{2}$O$_{5}$X$_{2}$ (X=Br or Cl) have attracted particular attention. These isostructural copper tellurates show intriguing ground state properties, with these magnetic compounds reportedly lying close to a quantum critical point\cite{lemmens,gros}. Previously, anomalies have been observed in the temperature dependence of magnetization and heat capacity measurements\cite{johnsson, lemmens} at $T^{Cl}_{N}$ =18.2 K and $T^{Br}_{N}$= 11.4 K for the Cl and Br compounds respectively. Recent neutron diffraction studies\cite{zaharko} have revealed that these transitions correspond to the onset of a similar incommensurate magnetic order. Interestingly, there is also a sharp drop in the magnetisation of both compounds at $T\sim$ 25 K, which has been attributed to the presence of a singlet-triplet spin-gap\cite{johnsson,lemmens}. This might suggest the possible coexistence of long range order with a singlet ground state. It is clear that there is a complicated interplay between the localised intra-tetrahedral interactions, which support spin-gapped behavior, and the inter-tetrahedral coupling that allows magnetic order to develop. The true nature of the groundstate in these systems remains a question for further elaboration. 

\begin{figure}[t]
\centering
\includegraphics[width = 8.5 cm]{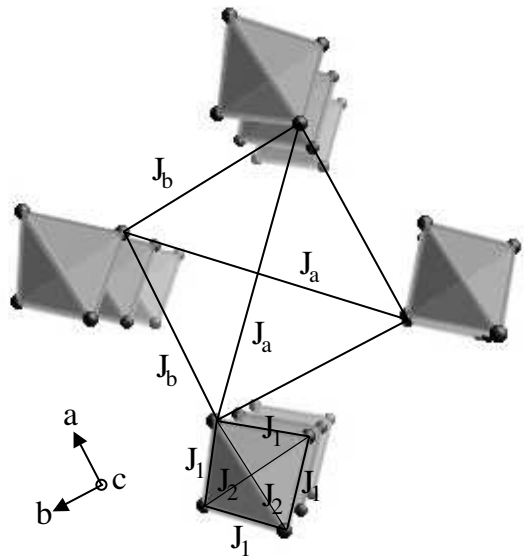}
\caption{Cu$^{2+}$ ions are located at the vertices of Cu$_4$O$_8$X$_4$ tetrahedra (O and X atoms not shown). The spin tetrahedra align in chains along the c-axis. Four exchange interaction paths are indicated; $J_1$ and $J_2$ are \textit{intra}-tetrahedral whilst $J_{a}$ and $J_{b}$ are \textit{inter}-tetrahedral.} 
\label{xtal}
\end{figure}

\par
The Cu$_{2}$Te$_{2}$O$_{5}$X$_{2}$ structure may be described in terms of Cu$_4$O$_8$X$_4$ tetrahedral clusters, with four spin-$1/2$ Cu$^{2+}$ ions situated at the vertices of the tetrahedra (see figure~\ref{xtal}). The clusters form a chain-like arrangement along the c-axis, separated from each other by Te and O atoms, with $\sim$3\% larger inter-chain separation for X = Br than for X = Cl. One can model the magnetic structure of Cu$_{2}$Te$_{2}$O$_{5}$X$_{2}$ as isolated tetrahedra by neglecting the inter-tetrahedral coupling and considering only nearest neighbor ($J_{1}$) and next nearest neighbor($J_{2}$) interactions, see figure~\ref{xtal}. The ground state of this model consists either of two singlet dimers or a quadrumer singlet involving all four Cu atoms, depending on the relative strength of $J_1$ and $J_2$. A good fit to magnetization data\cite{johnsson} gives $J_1$=$J_2$, in which case the system has a nonmagnetic singlet ground state, an excited triplet state and a singlet-triplet spin-gap of magnitude $J_1$=$J_2$=$J\sim$ 43 K  and 38.5 K for X = Br and Cl respectively. Despite the good fit, the validity of the independent tetrahedral model for this system has since been investigated, with further theoretical treatments\cite{brenig1,totsuka,gros,brenig,jensen,valenti,whangbo} looking at various inter-tetrahedral coupling configurations. Both Valenti \textit{et al.}\cite{valenti} and Whangbo \textit{et al.}\cite{whangbo} argue that the intercluster super-superexchange paths Cu-X$\cdot\cdot$X-Cu along the (110) direction and the (100) and (010) directions (denoted $J_{a}$ and $J_{b}$ respectively in figure~\ref{xtal}) are significant. Further work\cite{jensen,kotov} suggests the importance of antisymmetric Dzyaloshinsky-Moriya (DM) spin-spin interactions, which in a tetrahedral system can induce weak antiferromagnetic order from a singlet background. Experimentally, the excitation spectra of Cu$_{2}$Te$_{2}$O$_{5}$Br$_{2}$ has been probed by Raman spectroscopy\cite{lemmens,gros}, revealing evidence of both singlet-triplet and singlet-singlet excitations as well as the observation of a longitudinal magnon. However, without further experimental evidence it remains difficult to ascertain which exchange coupling configuration most closely describes Cu$_{2}$Te$_{2}$O$_{5}$Br$_{2}$ and Cu$_{2}$Te$_{2}$O$_{5}$Cl$_{2}$. In particular, in order to understand the nature and origin of the magnetic behavior in this system, it is important to measure the energy-momentum dispersion relation of the magnetic excitations. The experimental tool for doing this is neutron inelastic scattering (NIS), where the dispersion of the excitations and their intensity give information about the possible interactions and their relative strengths. In this paper we present NIS and neutron diffraction results from polycrystalline samples of both Cu$_{2}$Te$_{2}$O$_{5}$Cl$_{2}$ and Cu$_{2}$Te$_{2}$O$_{5}$Br$_{2}$.
\par
Polycrystalline samples of Cu$_{2}$Te$_{2}$O$_{5}$X$_{2}$ (X=Br, Cl) were prepared using the method described by Johnsson \textit{et al.}\cite{johnsson}. Magnetic susceptibility measurements were performed using a Quantum Design SQUID magnetometer, revealing only small
impurity-related Curie tails at low temperatures (below $\sim$ 5 K) corresponding to 0.2\% and 0.6\% free Cu$^{2+}$ impurity in the X=Br and X=Cl samples respectively. NIS measurements were carried out at the HET and MARI time of flight chopper spectrometers at the ISIS pulsed neutron facility, Rutherford Appleton Laboratory, UK. In addition, neutron diffraction measurements on the same polycrystalline samples were performed on the high flux D20 diffractometer at the ILL reactor, France, with a neutron wavelength of $\lambda$ = 2.4 \AA. 
\begin{figure}
\centering
\subfigure[X = Br]{\includegraphics[width = 8.5 cm]{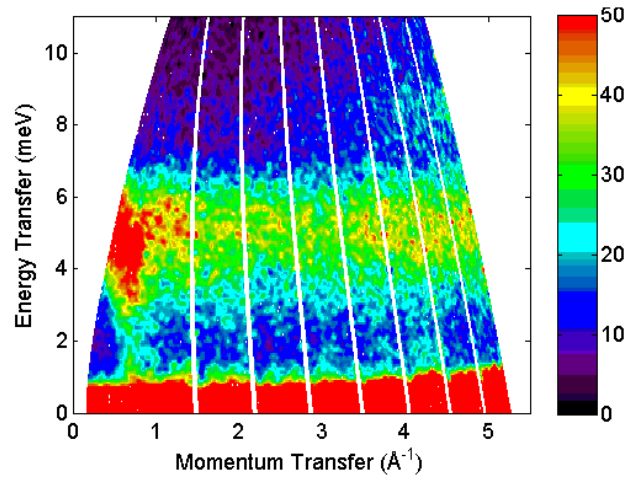}\label{2DBr}}
\qquad
\subfigure[X = Cl]{\includegraphics[width = 8.5 cm]{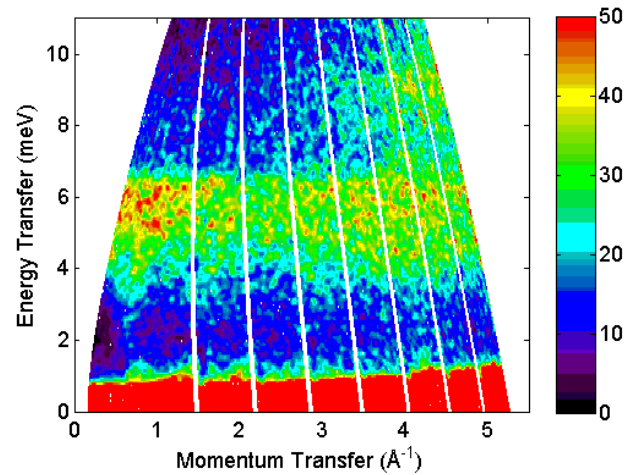}\label{2DCl}}
\caption{2D color map of the magnetic scattering in (a) Cu$_{2}$Te$_{2}$O$_{5}$Br$_{2}$  and (b) Cu$_{2}$Te$_{2}$O$_{5}$Cl$_{2}$ as a function of energy transfer($\hbar \omega$) and momentum transfer($Q$), obtained at 8 K, with incident energy 17 meV. The color scale denotes the scattering intensity ($S(|\mathbf{Q}|,\hbar \omega$), arb.~units).}
\end{figure}

\begin{figure}[t]
\centering
\includegraphics[width = 9.0 cm]{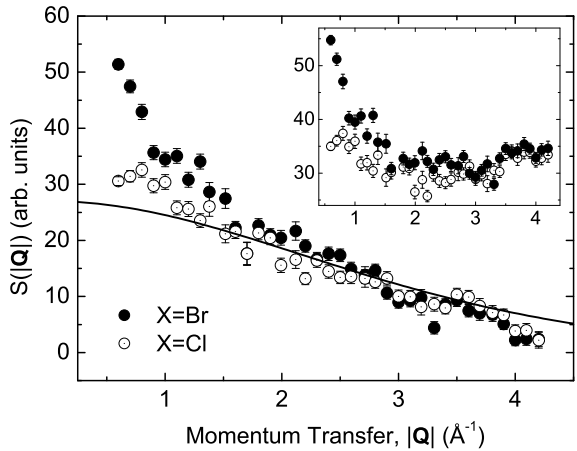}
\caption{Scattering intensity of Cu$_{2}$Te$_{2}$O$_{5}$Br$_{2}$  and Cu$_{2}$Te$_{2}$O$_{5}$Cl$_{2}$ as a function of momentum transfer, taken over an energy range 3-7 meV, at T = 8 K. The solid line represents the square of the Cu$^{2+}$ magnetic form factor~\cite{brown}. The inset shows the same data without corrections for multiple scattering and without the phonon contribution subtracted.}
\label{SQ}
\end{figure}

\begin{figure}
\centering
\subfigure[X = Br]{\includegraphics[width = 8.5 cm]{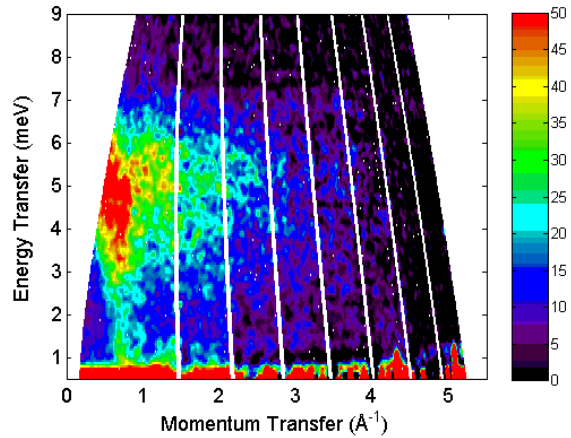}\label{2DBrcorr}}
\qquad
\subfigure[X = Cl]{\includegraphics[width = 8.5 cm]{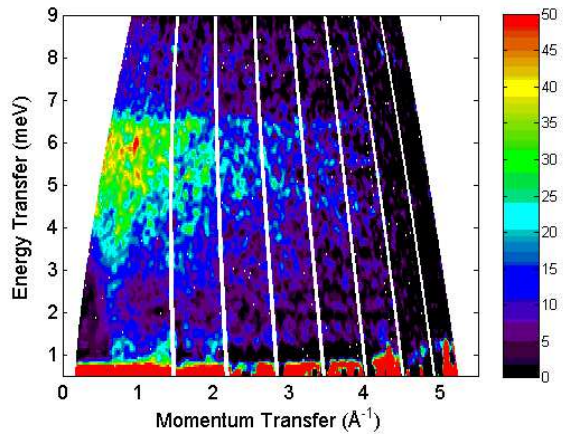}\label{2DClcorr}}
\caption{2D color map of the magnetic scattering in (a) Cu$_{2}$Te$_{2}$O$_{5}$Br$_{2}$  and (b) Cu$_{2}$Te$_{2}$O$_{5}$Cl$_{2}$ with corrections made for multiple scattering, and with the phonon contribution subtracted.}
\end{figure}

\begin{figure}[t]
\centering
\includegraphics[width = 8.5 cm]{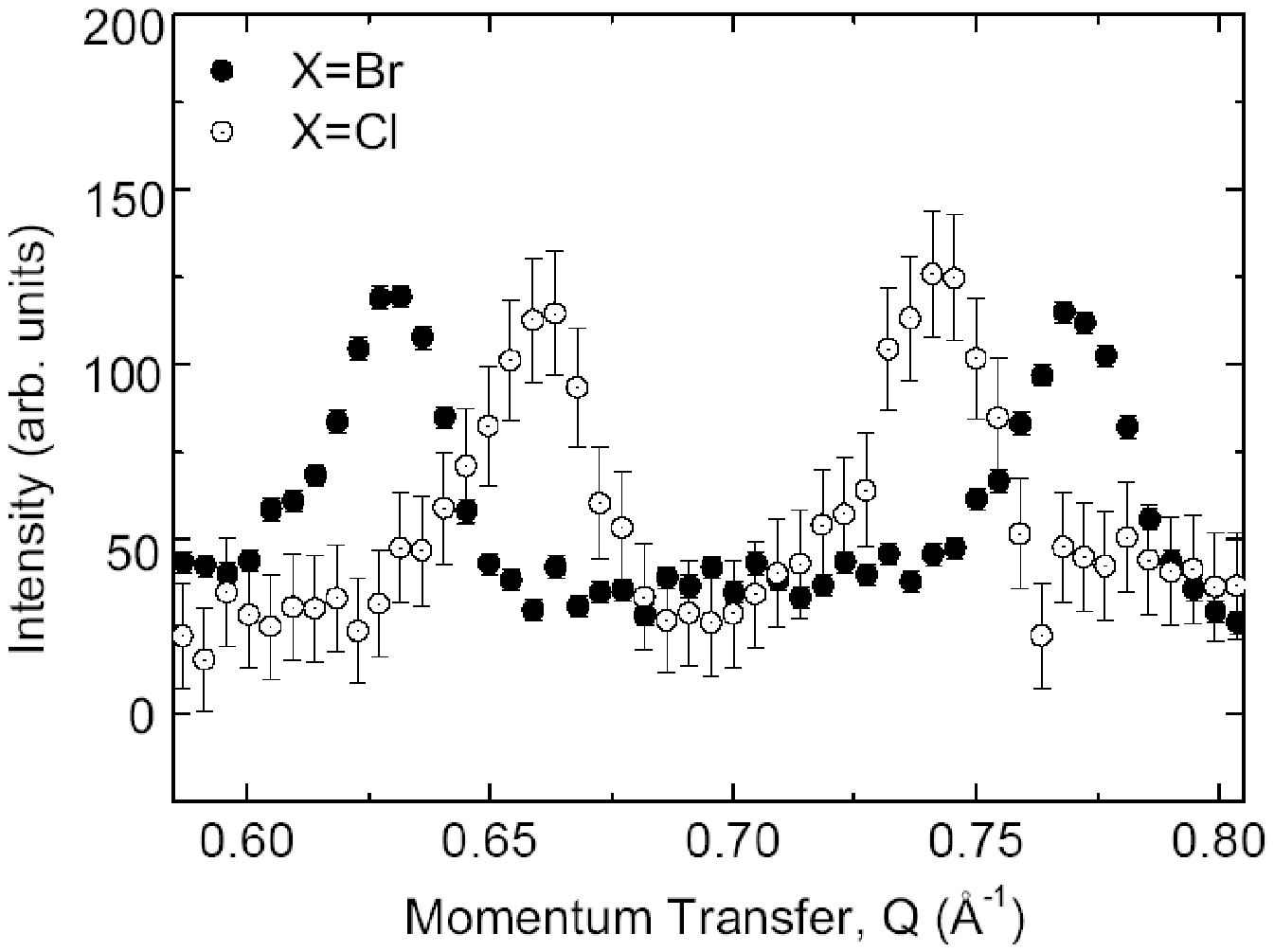}
\caption{Elastic neutron scattering of Cu$_2$Te$_2$O$_5$X$_2$ (X=Br,Cl), measured at the high flux D20 diffractometer, ILL. This is a difference plot, showing the high temperature scattering intensity (T = 25 K) subtracted from the low temperature scattering intensity (T = 2 K) as a function of momentum transfer.}
\label{D20}
\end{figure}

\par
Figures~\ref{2DBr} and ~\ref{2DCl} are 2D color plots of the neutron scattering data at 8 K with an incident energy of $E_i$ = 17 meV for Cu$_2$Te$_2$O$_5$Br$_2$ and Cu$_2$Te$_2$O$_5$Cl$_2$ respectively. The elastic line contains Bragg peaks, which can all be indexed on the basis of the crystal structure of Cu$_2$Te$_2$O$_5$X$_2$. Due to the polycrystalline nature of the samples, the scattering function $S(|\mathbf{Q}|,\hbar \omega)$ in these measurements is the powder average of the spin-spin correlation function $S(\textbf{Q},\hbar \omega)$, and is not sensitive to directions in momentum space. In our data, $S(|\mathbf{Q}|,\hbar \omega)$ shows a clear band of strong intensity centred about an energy of approximately 5.5 meV and 6 meV for X=Br and X=Cl respectively. The magnetic character of these peaks in $S(|\mathbf{Q}|,\hbar \omega)$ is seen from the decreasing intensity with increasing momentum transfer in the low $|\mathbf{Q}|$ region ($|\mathbf{Q}|<$ 3 \AA$^{-1}$). This is more clearly illustrated in figure~\ref{SQ}, which depicts $S(|\mathbf{Q}|,\hbar \omega)$ versus momentum transfer taken for the energy transfer range 3 - 7 meV. It is clear from figures~\ref{2DBr}, \ref{2DCl} and~\ref{SQ} that the bands of intensity centered at $\sim$ 5.5 meV (X=Br) and $\sim$ 6 meV (X=Cl) extend in to the high $|\mathbf{Q}|$ region, where one expects vanishing intensity of the magnetic excitation in both compounds due to the magnetic form factor.  At $|\mathbf{Q}|$ = 3 \AA$^{-1}$ the Cu$^{2+}$ form factor is $\sim$46$\%$ of its value at $|\mathbf{Q}|$ = 0.5 \AA$^{-1}$, and by $|\mathbf{Q}|$ = 5 \AA$^{-1}$ it is $\sim$14$\%$. However, in our data the scattering intensity levels off to an almost constant value between 2 \AA$^{-1}<|\mathbf{Q}|<$ 3 \AA$^{-1}$ and begins to increase above $|\mathbf{Q}| \sim$ 3 \AA$^{-1}$. This indicates the dominance of vibrational modes above $\sim$ 4 \AA$^{-1}$ (where a second phonon mode at $\sim$ 9.5 meV is also present in both samples). On this basis, we have assumed a purely vibrational contribution to the inelastic scattering detected in the high $|\mathbf{Q}|$ detector bank, and from this we have calculated the phonon contribution as well as multiple scattering contribution as a function of $|\mathbf{Q}|$ and $\hbar \omega$ using the DISCUS simulation program\cite{discus}. We have subtracted the phonon and multiple scattering contributions in order to obtain a purely magnetic response from our data, shown in figures~\ref{2DBrcorr} and~\ref{2DClcorr} as 2D color plots of $S(|\mathbf{Q}|,\hbar \omega)$. Figure~\ref{SQ} shows $S(|\mathbf{Q}|,\hbar \omega)$ versus momentum transfer for the corrected data, with the Cu$^{2+}$ magnetic form factor~\cite{brown} also displayed (solid line). For both samples the $|\mathbf{Q}|$-dependence of the scattering above $\sim$1.5 \AA$^{-1}$ is close to that expected from the Cu$^{2+}$ form factor, but at low $|\mathbf{Q}|$ ($<$ 1.5 \AA$^{-1}$) the scattering intensity shows a clear deviation from the single-ion Cu$^{2+}$ form factor, particularly in the bromide. This suggests that there is a structure factor effect in addition to the form factor, which arises from a larger sized scattering entity, the nature of which (e.g. tetrahedral or square planar arrangement of Cu$^{2+}$ ions) depends upon the relative strength of the exchange interactions present. The presence of larger scattering entities would manifest itself in the superposition of $|\mathbf{Q}|$-dependent oscillations about the single-ion form factor. This would be most dominant at low $|\mathbf{Q}|$ and dampen with increasing $|\mathbf{Q}|$. The large increase in intensity at low $|\mathbf{Q}|$ for the bromide may suggest the dominance of inter-tetrahedral exchange interactions on the length scale of $J_a$ and $J_b$ (see figure~\ref{xtal}).

\par
Figure~\ref{D20} shows results from neutron diffraction measurements of polycrystalline  Cu$_{2}$Te$_{2}$O$_{5}$X$_{2}$ (X=Cl,Br), depicting the difference in the scattering intensity above (25 K) and below (2 K) the transition temperature ($S_{2K}$ - $S_{25K}$) as a function of $|\mathbf{Q}|$. The magnetic order in both systems is incommensurate, with modulation vectors $\mathbf{k^{Br}}$= (0.19, 0.36, 0.5) and $\mathbf{k^{Cl}}$= (0.15, 0.40, 0.5) for X=Br and Cl respectively. The ground state magnetic order below $T_N$ is rather complicated, and cannot be determined solely from polycrystalline measurements. Single crystal measurements of the X=Cl sample reported by Zaharko \textit{et al.}\cite{zaharko} show the magnetic structure to involve multiple helices, and the crystals to possess more than one domain. Single crystal measurements of the X=Br sample are required to fully understand its magnetic structure. However, we note that magnetic peaks arising from the incommensurate magnetic order are observed at $|\mathbf{Q}|$= 0.63 and 0.77 \AA$^{-1}$ for X=Br, and $|\mathbf{Q}|$= 0.66 and 0.74 \AA$^{-1}$ for X=Cl respectively, as shown in figure~\ref{D20}. As we discuss later, our NIS measurements on both compounds reveal a dispersive mode with the minimum energy occurring at the same position in $|\mathbf{Q}|$ as these incommensurate Bragg peaks.

\begin{figure}[b]
\centering
\includegraphics[width = 9.0 cm]{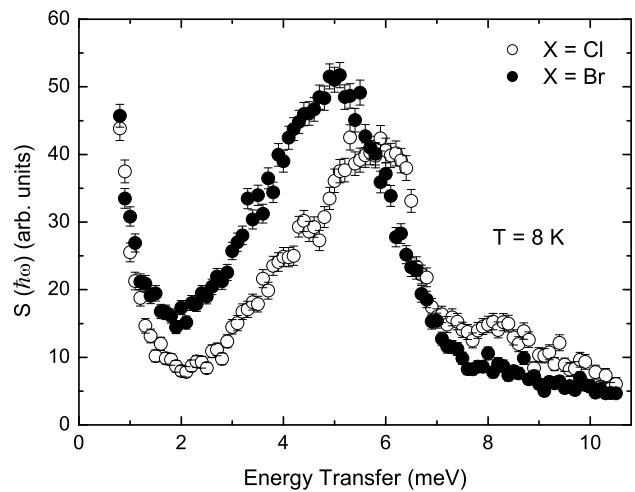}
\caption{Scattering intensity versus energy transfer for X = Br and Cl, summed over the low angle detector bank (scattering angle 3-13$^o$, which corresponds approximately to 0.5 \AA$^{-1}<|\mathbf{Q}|<$1.4 \AA$^{-1}$), taken at T = 8 K.} 
\label{SE}
\end{figure}

\par
In the 2D plots for inelastic scattering of Cu$_{2}$Te$_{2}$O$_{5}$X$_{2}$ (figures~\ref{2DBrcorr} and~\ref{2DBrcorr}), two components to the magnetic excitations are observed; a flat, constant energy ($\sim$ 5.5 meV for X=Br and $\sim$ 6 meV for X=Cl) component that falls in intensity with increasing $Q$ as previously discussed, and a narrow, dispersive band of intensity centered at $|\mathbf{Q}| \sim$ 0.7 \AA$^{-1}$ in the energy range around 1 to 3 meV. The dispersive component is particularly clear in the bromide, where it stretches toward the magnetic Bragg peaks observed in elastic measurements (see figure~\ref{D20}), suggesting that the dispersive excitation is supported by the incommensurate magnetic order. The dispersive component in the chloride is relatively weaker, and there is a clear energy gap between the magnetic Bragg peaks and the excitation.

\par
\begin{figure}[t]
\centering
\subfigure[X = Br]{\includegraphics[width = 9.0 cm]{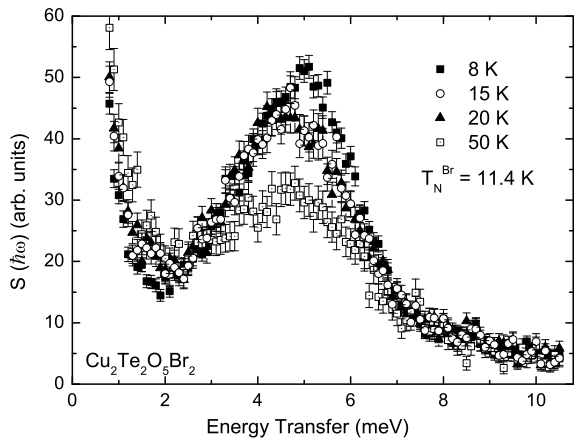}\label{SETBr}}
\qquad
\subfigure[X = Cl]{\includegraphics[width = 9.0 cm]{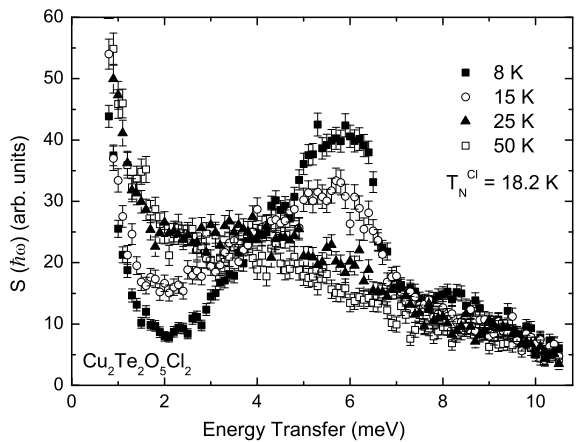}\label{SETCl}}
\caption{Scattering intensity versus energy transfer at 8 K, 15 K and 50 K for Cu$_{2}$Te$_{2}$O$_{5}$Br$_{2}$ and Cu$_{2}$Te$_{2}$O$_{5}$Cl$_{2}$. The data is summed over the low angle detector bank (scattering angle 3-13$^o$, which corresponds approximately to 0.5 \AA$^{-1}<|\mathbf{Q}|<$1.4 \AA$^{-1}$). }
\end{figure}

\par
The energy dependence of the magnetic peak in each compound at $T$ = 8 K are shown in figure~\ref{SE}, which is an $S(|\mathbf{Q}|,\hbar \omega)$ versus $\hbar \omega$ plot summed over the low angle detector bank (scattering angle 3 - 13$^o$). The corresponding $|\mathbf{Q}|$ range of the summation is energy dependent but does not vary significantly over the energy range of interest, and is approximately 0.5 \AA$^{-1}<|\mathbf{Q}|<$1.4 \AA$^{-1}$. Summing over an entire detector bank rather than a fixed $|\mathbf{Q}|$ range enables us to avoid problems with interpolation and extrapolation of the data. For both samples the excitation has a width that it is not resolution limited (the instrumental resolution is $\sim$ 0.5 meV at 5 meV), due either to dispersion or a lifetime effect. In the bromide sample there is little structure in the excitation and only one peak is resolved, which is Gaussian in shape to a reasonable approximation. In contrast, the chloride excitation peak is asymmetric, with a sharp fall on the high energy side, and appears to be made up of constituent peaks. 

\par
Now let us look at the temperature dependence (at 8 K, 15 K, 20 K (or 25 K, X=Cl) and 50 K) of the $S(|\mathbf{Q}|,\hbar \omega)$ versus $\hbar \omega$ data, shown for the Br and Cl compounds in figures~\ref{SETBr} and ~\ref{SETCl} respectively. The two compounds show a striking contrast in their temperature dependence. Firstly, for Cu$_{2}$Te$_{2}$O$_{5}$Br$_{2}$ the integrated intensity of the peak decreases smoothly with increasing temperature, showing no change in lineshape across the transition temperature $T^{Br}_{N}\sim$ 11.4 K. Both the flat, constant energy component and the dispersive component are present at all temperatures, falling uniformly in intensity with increasing temperature. A very different temperature dependence is observed in the X=Cl compound, in which the centre of mass of the scattering moves to lower energy with increasing temperature. Initially, a fall in the intensity of the 6 meV peak is observed when the temperature increases from 8 K to 15 K, accompanied by an increase in the intensity around 2.5 meV. However, above the transition temperature $T^{Cl}_{N}$ =18.2 K, the inelastic structure is replaced by a quasi elastic lineshape from a diffusive response.


\par Our present work has investigated the nature of the magnetic excitations in the spin tetrahedral compounds Cu$_2$Te$_2$O$_5$X$_2$ (X = Br, Cl). Firstly, let us consider that the system consists of independent tetrahedra of Cu$^{2+}$ ions. In this model, the excitations are dispersionless, and the $|\mathbf{Q}|$-dependence follows the structure factor of a single tetrahedron, characterised by exchange couplings on intra-tetrahedral distances. In our data, the dispersionless component of the excitations centered at $\sim$ 5.5 meV (X=Br) and 6 meV (X=Cl) would correspond, in the isolated tetrahedral model, to singlet-triplet spin-gaps of $\Delta_{Br}$ = 5.5 meV and $\Delta_{Cl}$ = 6 meV.
\par
If one takes inter-tetrahedral coupling into account, the magnetic units of interest may no longer be tetrahedra but, for example, isolated plaquettes such as the square planar arrangement displayed in figure~\ref{xtal}, which is mediated by exchange couplings that act over longer inter-tetrahedral distances. If these are non-interacting, they also give rise to dispersionless excitations, but with a $|\mathbf{Q}|$-dependence arising from the structure factor of an inter-tetrahedral exchange configuration. The difference observed between the $|\mathbf{Q}|$-dependence of the scattering functions for Cu$_2$Te$_2$O$_5$Br$_2$ and Cu$_2$Te$_2$O$_5$Cl$_2$ is indicative of different underlying exchange configurations. In the bromide, the steep fall in scattering intensity with increasing $|\mathbf{Q}|$ in the low $|\mathbf{Q}|$ region compared to the Cu$^{2+}$ form factor suggests the importance of exchange interactions on an inter-tetrahedral length scale. The presence of inter-tetrahedral couplings also produces a splitting of the energy states of the isolated tetrahedral model, thereby allowing several excitation modes, and the presence of multiple peaks. As the relative population of the energy states develop with temperature, so would the relative intensity of the corresponding excitation peaks. In our data we are unable to resolve multiple peaks, and are therefore unable to observe an effect such as this.
\par
In our data we see clear evidence of dispersive magnetic excitations, which are associated with the magnetic Bragg peaks that arise from the incommensurate order below $T_N$. The dispersion of excitations in a conventional antiferromagnet are approximately linear at the Brillouin zone (BZ) center, but flatten off at the zone boundary. This behavior is observed in the low $|\mathbf{Q}|$ region of our data ($|\mathbf{Q}| < $1.5 \AA$^{-1}$), but does not repeat itself periodically as we move to larger $|\mathbf{Q}|$. This is due to the fact that this is a polycrystalline measurement in which the scattering intensity at a given $|\mathbf{Q}|$ is the average intensity over a sphere of radius $|\mathbf{Q}|$ in reciprocal space. At low $|\mathbf{Q}|$, the scattering sphere only encompasses a small number of BZs and therefore the strongly dispersive center of the zone is sampled effectively. At large $|\mathbf{Q}|$ the scattering intensity is averaged over a sphere covering many BZs, and therefore approaches a density of states for the magnetic excitations, which the flat component at the zone boundary dominates. Therefore, in our data we see only the flat component of the excitations at high $|\mathbf{Q}|$, and the dispersive feature appears only at low $|\mathbf{Q}|$. In order to fully understand the dispersive excitations in this system, single crystal inelastic measurements are required.
\par
The difference in the observed temperature dependence of these two materials is particularly interesting. Cu$_2$Te$_2$O$_5$Cl$_2$ shows a structural inelastic response below the magnetic ordering transition, $T^{Cl}_{N}$, and a diffusive response corresponding to short range correlations in the paramagnetic state. Cu$_2$Te$_2$O$_5$Br$_2$ on the other hand, shows an unusual temperature dependence with respect to the transition at $T^{Br}_{N}$. In fact, the excitation spectra shows no significant change at $T^{Br}_{N}$, but continues to fall uniformly with increasing temperature, similar to what might be expected from a two-level system such as a simple singlet-triplet configuration. However, the dispersive component of the excitation is present well above $T^{Br}_{N}$, and falls in intensity at the same rate as the flat component with increasing temperature. 
A tendency of the system to short range order, or the build up of low dimensional order well above $T^{Br}_{N}$ may provide a mechanism by which the dispersive excitations are supported above $T^{Br}_{N}$, thus explaining the temperature dependence observed in the bromide. This may also explain the small anomaly observed in heat capacity measurements~\cite{lemmens} at $T^{Br}_{N}$, as the presence of low dimensional correlations at higher temperatures results in less entropy being associated with the transition to 3D long range order. Despite the similarity of the incommensurate magnetic order of these materials below $T^{Br,Cl}_{N}$, the nature of their transition are markedly different, as supported by heat capacity~\cite{lemmens} and thermal conductivity~\cite{sologubenko} measurements in addition to the results presented here.

\par
In conclusion, magnetic excitations with a dispersive component have been directly observed in inelastic neutron scattering measurements of the spin tetrahedral Cu$_2$Te$_2$O$_5$Br$_2$ (X=Br,Cl) materials. The chloride supports an excitation spectra which softens toward the ordering temperature, whilst the bromide shows an unusual temperature dependence indicating the possibility of low dimensional and/or short range correlations above $T^{Br}_{N}$. We have shown that there are large differences in the dynamic behavior of these two compounds, despite the similarity of their static magnetic order below $T^{Br,Cl}_{N}$. However, the possible scenarios will only be tested reliably when large single crystal mosaics become available, allowing a clear measurement of the dispersive magnetic excitations to be preformed.
\par
The authors would like to thank T. Orton for technical support, and acknowledge the financial support of the UK Engineering and Physical Sciences Research Council. 

\par

\newpage


\begin{thebibliography}{99}

\bibitem{lemmens} P. Lemmens, K.-Y. Choi, E. E. Kaul, C. Geibel,K. Becker, W. Brenig, R. Valent\'i, C. Gros, M. Johnsson, P. Millet, and F. Mila, Phys. Rev. Lett. \textbf{87}, 227201 (2001).

\bibitem{gros} C. Gros, P. Lemmens, M. Vojta, R.Valent\'i, K.-Y. Choi, H. Kageyama, Z. Hiroi, N.V. Mushnikov, T. Goto, M. Johnsson, and P. Millet, Phys. Rev. B \textbf{67}, 174405 (2003).

\bibitem{johnsson} M. Johnsson, K. W. Tornroos, F. Mila, and P. Millet, Chem. Mater. \textbf{12}, 2853 (2000).

\bibitem{zaharko} O. Zaharko, A. Daoud-Aladine, S. Streule, J. Mesot, P.-J. Brown, and H. Berger, Phys. Rev. Lett. \textbf{93}, 217206 (2004).

\bibitem{brenig1} W. Brenig and K.W. Becker, Phys. Rev. B \textbf{64}, 214413 (2001). 

\bibitem{totsuka} K. Totsuka and H.-J. Mikeska, Phys. Rev. B \textbf{66}, 054435 (2002).

\bibitem{brenig} W. Brenig, Phys. Rev. B \textbf{67}, 064402(2003).

\bibitem{jensen} J. Jensen, P. Lemmens, and C. Gros, Europhys. Lett. \textbf{64}, 689 (2003).

\bibitem{valenti} R. Valent\'i, T. Saha-Dasgupta, C.Gros, and H. Rosner, Phys. Rev. B \textbf{67}, 245110 (2003).

\bibitem{whangbo} M.-H. Whangbo, H.-J. Koo, and D. Dai, Inorg. Chem \textbf{42}, 3898 (2003).

\bibitem{kotov} V.N. Kotov, M.E. Zhitomirsky, M. Elhajal, and F. Mila, Phys. Rev. B \textbf{70}, 214401 (2004).

\bibitem{discus} M.W. Johnson, UKAEA Harwell Report AERE-R7682 (1974).

\bibitem{brown} P.J. Brown, in \textit{International tables for Crystallography}, edited by A.J.C. Wilson (Kluwer Academic, Dordrecht, 1992), Vol. C, p. 391.

\bibitem{sologubenko} A.V. Sologubenko, R. Dell'Amore, H.R. Ott, and P. Millet, Euro. Phys. Journ. B \textbf{42}, 549 (2004).
 
\end{thebibliography}
\end{document}